\documentclass[twocolumn]{aastex62}
\usepackage[utf8]{inputenc}
\usepackage{amsmath}
\usepackage{graphicx}
\usepackage{savesym}
\savesymbol{tablenum}
\usepackage{siunitx}
\restoresymbol{SIX}{tablenum}

\makeatletter
\newcommand{\mr}[1]{\mathrm{#1}}
\newcommand{\paren}{\@ifstar\@@paren\@paren}
\newcommand{\@paren}[2][]{\if\relax\detokenize{#1}\relax\left( #2 \right)\else\csname#1l\endcsname( #2 \csname#1r\endcsname)\fi}
\newcommand{\@@paren}[2][]{( #2 )}
\newcommand{\abs}{\@ifstar\@@abs\@abs}
\newcommand{\@abs}[2][]{\if\relax\detokenize{#1}\relax\left\lvert #2 \right\rvert\else\csname#1l\endcsname\lvert #2 \csname#1r\endcsname\rvert\fi}
\newcommand{\@@abs}[2][]{\lvert #2 \rvert}
\newcommand{\dd}[2][]{\mathop{}\mkern-4mud^{#1}\mkern-0.5mu#2}
\newcommand{\der}[3][]{\frac{\dd[#1]{#2}}{\dd{#3}^{#1}}}
\newcommand{\bracket}{\@ifstar\@@bracket\@bracket}
\newcommand{\@bracket}[2][]{\if\relax\detokenize{#1}\relax\left\langle #2 \right\rangle\else\csname#1l\endcsname\langle #2 \csname#1r\endcsname\rangle\fi}
\newcommand{\@@bracket}[2][]{\langle #2 \rangle}
\makeatother
\newcommand{\cross}{\times}
\newcommand{\eps}{\varepsilon}
\newcommand{\ee}{e}
\let\of\paren
\let\norm\abs
\let\vec\mathbf
\newcommand{\vhat}[1]{\mathbf{\hat{#1}}}
\usepackage{layouts}

\begin{document}

\title{Detecting gravitational scattering of interstellar objects using pulsar timing}
\author{Ross J. Jennings}
\affiliation{Department of Astronomy, Cornell University, Ithaca, NY 14853, USA}
\author{James M. Cordes}
\affiliation{Department of Astronomy, Cornell University, Ithaca, NY 14853, USA}
\affiliation{Cornell Center for Astrophysics and Planetary Science, Cornell University, Ithaca, NY 14853, USA}
\author{Shami Chatterjee}
\affiliation{Department of Astronomy, Cornell University, Ithaca, NY 14853, USA}
\affiliation{Cornell Center for Astrophysics and Planetary Science, Cornell University, Ithaca, NY 14853, USA}

\begin{abstract}
 Gravitational scattering events, in which the path of an interstellar object is deflected by a pulsar or the solar system, give rise to reflex motion which can potentially be detected using pulsar timing.  We determine the form of the timing signal expected from a gravitational scattering event, which is ramp-like and resembles the signal produced by a glitch or a gravitational wave burst with memory (BWM), and investigate the prospects for detecting such a signal using a pulsar timing array. The level of timing precision currently achieved for some millisecond pulsars makes it possible to detect objects as small as $10^{-10}$\,$M_\sun$, less than the mass of the dwarf planet Ceres, at impact parameters as large as $\SI{1}{au}$. The signals produced by gravitational scattering could provide independent constraints on models of dark matter involving asteroid-mass objects or subhalos, and should be considered as potential false positives in searches for BWMs.
\end{abstract}

\section{Introduction}

If a compact astrophysical object were to pass near enough to a pulsar or to the solar system to interact gravitationally, it would slightly alter the motion of the pulsar relative to the solar system barycenter, potentially producing a detectable effect on the pulse times of arrival (TOAs). The North American Nanohertz Observatory for Gravitational Waves (NANOGrav; \citealt{11yr-timing}) and other pulsar timing arrays (PTAs;  \citealt{dcl+16,hobbs13}) have collected hundreds of pulsar-years of high-precision (sub-microsecond) pulsar timing data as part of their efforts to detect gravitational waves. 
The various regional PTAs collaborate to form the international pulsar timing array (IPTA; \citealt{ipta10}), which occasionally releases combined data sets (e.g. \citealt{ipta-dr1,ipta-dr2}).
The qualities of PTAs that make them well suited for detecting gravitational waves -- namely, the precision of arrival time measurements and the long spans of time over which they may be collected -- also make them highly sensitive to perturbations of this type. 

An interstellar object (ISO) gives rise to a gravitational scattering signal in pulsar timing data if it passes close enough to a pulsar or to the solar system. One important category of ISOs consists of free-floating planets and smaller asteroid- and comet-like bodies. Almost certainly, a large number of these bodies exist in interstellar space, since the planet formation process is thought to result in the ejection of large numbers of rocky bodies, ranging in size from planetesimals to fully-formed planets \citep{cm03}. Some such bodies have even been directly observed: 
In particular, the asteroid 1I/`Oumuamua \citep{mwm+17} and the comet 2I/Borisov \citep{gpd+19} have both been identified as interstellar in origin. To produce a detectable timing signal, an asteroid-like ISO would have to be a few hundred kilometers in diameter and pass within a few au of a pulsar. This is significantly larger than either `Oumuamua or Borisov, but smaller than the largest objects in the asteroid belt.

Some theories of dark matter predict that at least a fraction of it is composed of massive ISOs, such as primordial black holes (PBHs) or subhalos. As a result, several searches for ISOs have been carried out with the goal of understanding the nature of dark matter. For the most part, these have involved microlensing surveys, which are sensitive to any object with mass along the line of sight to a star. The pioneering MACHO~\citep{macho}, EROS~\citep{eros}, and OGLE~\citep{ogle} surveys largely ruled out the possibility that the majority of dark matter could consist of objects between $10^{-8}$ and $100\,M_\sun$, but did not significantly constrain objects smaller than $10^{-8}\,M_\sun$ (approximately $20$ times the mass of Ceres). Subsequently, attention has focused on primordial black holes with masses between $10^{-11}$ and $10^{-8}\,M_\sun$ as dark matter candidates \citep{cks16}. The strongest limits on PBHs and other relatively compact objects in this mass range come from a recent microlensing survey of the Andromeda galaxy using the Hyper Suprime Cam (HSC) on the 8.2-meter Subaru telescope \citep{nty+19}. The implications of these constraints for the expected event rate are discussed further in Section~\ref{sec:freq}.

A PBH or similar dark matter constituent would interact with a pulsar in a manner indistinguishable from a baryonic ISO of equal mass. If they constituted a significant fraction of dark matter by mass, PBHs would also be much more numerous than baryonic ISOs of the same mass. Because of this, pulsar timing has been proposed as a means of detecting PBHs. Some proposals have focused on the Shapiro delay signal caused by objects along the line of sight to the pulsar \citep{shf07,cls16}, but others \citep{sc07,ks12,ko18,drtz19} have discussed the Doppler signal, which is more closely related to the timing signal described here.

Detectable encounters with ISOs are likely to be rare, but examining a large volume of data makes their observation more likely. In this respect, it is important to distinguish between encounters with a pulsar and encounters with the solar system. By analogy with the terminology used for gravitational wave signals, we call the former ``pulsar-term'' scattering events and the latter ``Earth-term'' events. If many pulsars are observed, the two scenarios give complementary sensitivities --- since Earth-term events affect all pulsars, it is possible to detect weaker events by cross-correlating the signals, but objects are more likely to pass near one of the pulsars in a PTA simply because there are more pulsars. This means that Earth-term events are more detectable for large populations of very small objects, but pulsar-term events are more detectable for small populations of larger objects.


In what follows, we calculate the shape of the timing signal expected from a gravitational scattering event in detail, and use the results to assess the circumstances under which such events may be detectable. In Section~\ref{sec:detector}, we discuss the factors which limit the precision of pulsar timing measurements. In Section~\ref{sec:deltatau}, we derive the expected timing signal. In Section~\ref{sec:pulsar-encounters}, we discuss the conditions under which pulsar-term events may be detected. Section~\ref{sec:solar-system} describes the corresponding conditions for Earth-term events. In Section~\ref{sec:freq}, we review methods of estimating the number density of potential perturbers in the galaxy, and speculate as to the frequency of detectable events. Finally, in Section~\ref{sec:summary}, we summarize our results and draw conclusions.

\section{Pulsar timing precision}\label{sec:detector}


Contemporary pulsar timing methods rely on the fact that every pulsar has a characteristic average pulse shape which is stable on time scales of decades. Arrival times are determined by comparing observed pulse shapes with the template pulse shape using a Fourier-domain matched filtering algorithm \citep{taylor92}. Generally, averages with $N\gg1$ pulses are used, with typical values of $N$ being between $10^5$ and $10^6$.

The minimum uncertainty in estimating an arrival time comes from noise in the pulse profile measurement introduced by the receiver (radiometer noise). Computing TOAs using a matched filtering algorithm minimizes this contribution to the arrival time error for a given level of radiometer noise. In the absence of other sources of error, the uncertainty in arrival times computed using matched filtering is given by \citep{cordes13,lcc+16}:
\begin{equation}\label{eqn:sigma-mf}
    \sigma_{\mr{MF}}=\frac{\sqrt{P\mkern1muW_{\mr{eff}}}}{S\sqrt{N_\phi}}.
\end{equation}
Here $P$ is the pulse period, $S$ is the signal-to-noise ratio (SNR), $N_\phi$ is the number of samples in pulse phase, and $W_{\mr{eff}}$ is an effective pulse width, given by
\begin{equation}
    W_{\mr{eff}}=\paren[bigg]{\int_0^P U'(t)^2\dd t}^{-1},
\end{equation}
where $U\of{t}$ is the pulse shape (normalized to unit maximum). 

In practice, TOA errors are larger than equation~(\ref{eqn:sigma-mf}) would predict, because other effects contribute. These include pulse jitter from the motion of emission regions in pulsar magnetospheres; dispersion and scintillation, caused by propagation of the signal through the ionized interstellar medium; and spin noise, caused by interactions between the neutron star crust and its magnetosphere and superfluid interior. Of these, pulse jitter is uncorrelated in time, but scintillation, dispersion measure variations, and spin noise are generally correlated, producing gradual but random drifts in pulse times of arrival. Spin noise has a ``red'' power spectrum with most of the power concentrated at low frequencies. By contrast, radiometer and jitter noise have a ``white'' power spectrum, contributing approximately equal power at all frequencies.

Broadly speaking, pulsars can be divided into two categories: canonical pulsars (CPs), which are relatively young and tend to have periods of order one second and surface magnetic fields of order $10^{12}$ gauss; and millisecond pulsars (MSPs), which are much older and are thought to have been spun up by accreting matter from a companion. MSPs have significantly shorter periods (a few milliseconds) and weaker magnetic fields (of order $10^8$ gauss) than canonical pulsars. Additionally, MSPs spin down much more slowly and have lower levels of spin noise (by a factor of around $10^6$) than CPs \citep{sc10,lcc+17,psj+19}. In part because of their short periods, pulses from MSPs can also be localized more precisely (cf. equation~\ref{eqn:sigma-mf}). For these reasons, pulsar timing arrays, which require very high-precision timing, almost exclusively time MSPs.

For many MSPs, TOA precision significantly better than \SI{1}{\micro s} is already routinely achieved. For example, 38 of the 45 pulsars included in the NANOGrav 11-year data set~\citep{11yr-timing} had a median TOA uncertainty in L-band (1--2\,\si{GHz}) observations which was below \SI{1}{\micro s}, with the best-timed pulsar, PSR B1937+21, having a median uncertainty of a mere \SI{12}{ns}. The overall median TOA uncertainty in the data set is \SI{319}{ns}. A detailed breakdown of per-pulsar contributions to excess noise in a previous NANOGrav data release, the 9-year data set, is given by~\citet{lcc+17}.

Because of its red spectrum, spin noise presents an additional challenge in searches for very low-frequency and quasi-static effects. While not all MSPs possess detectable levels of red spin noise, many do. We defer a full analysis of the effects of red spin noise on the detectability of scattering events to future work, but note that its effects are likely to be significant for events with large impact parameter, $b$, for which most of the signal power is concentrated at low frequencies (see Section~\ref{sec:deltatau} below).

The demands on timing precision for gravitational wave detection require attention to all processes that contribute more than $\sim$\SI{10}{ns} RMS to timing residuals. \citet{scm+13} considered the effects of asteroid belts around isolated MSPs and in particular determined that the nonstationary timing residuals for B1937+21 are not inconsistent with an asteroid belt interpretation.

\section{TOA perturbations}\label{sec:deltatau}

The detectable effect of a perturbing object on measured TOAs is a result of the reflex motion of the pulsar or the solar system barycenter. For concreteness, we assume the perturbing object passes near the pulsar, with the understanding that the roles of the pulsar and the solar system barycenter can be interchanged if the perturber passes through the solar system. We also treat the perturber as a point mass, and assume that general relativistic effects and  non-gravitational forces (such as the Yarkovsky effect, e.g. \citealt{rubincam98}) can be ignored. A potential complicating factor is the fact that many MSPs are found in binary systems with white dwarfs, with typical separations between \num{0.01} and \SI{0.6}{au}. In this case, we are concerned with the reflex motion of the center of mass of the pulsar system. If the perturber comes closer than a few times the binary separation, three-body interactions, the effects of which are beyond the scope of this paper, may become important.

\begin{figure}
    \centering
    \includegraphics[width=\linewidth]{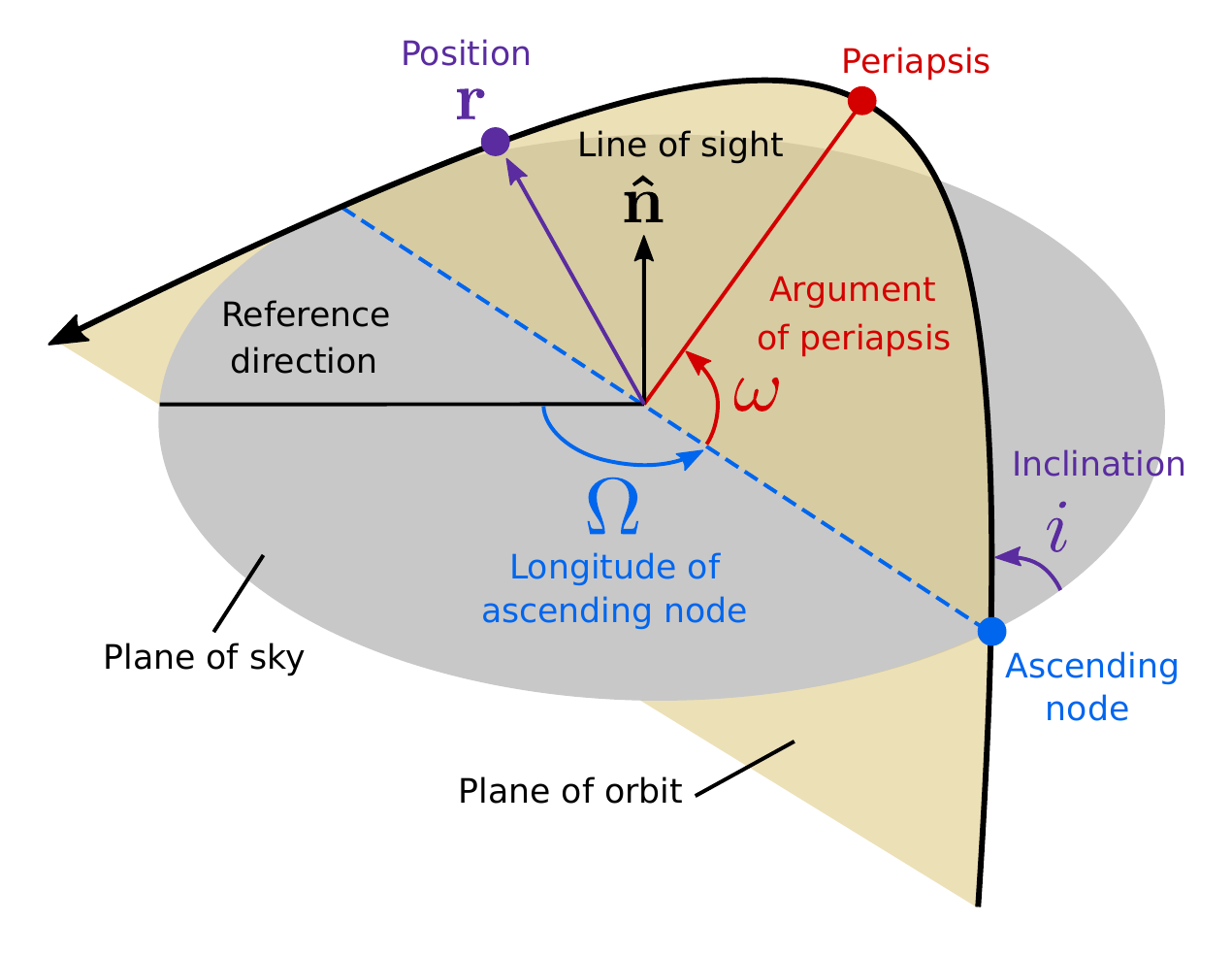}
    \caption{Geometry of a hyperbolic orbit, showing the three angles necessary to specify its orientation: the inclination, $i$, the argument of the periapsis, $\omega$, and the longitude of the ascending node, $\Omega$. Also shown are the position vector $\vec{r}$ of the orbiting body (here, the perturbing ISO) and the line of sight direction $\vec{\hat{n}}$.} The solid red line is the line of apsides, and the dashed blue line is the line of nodes.
    \label{fig:orbit-diagram}
\end{figure}

The position, $\vec{r}$, of the perturber relative to the center of mass of the pulsar-perturber system follows a hyperbolic trajectory, parameterized as
\begin{equation}\label{eqn:hyp}
\vec{r}\of{H}=\frac{b\mkern1mu\paren{e-\cosh{H}}}{\sqrt{e^2-1}}\mkern1mu\vhat{x}+b\mkern1mu\sinh{H}\mkern1mu\vhat{y}
\end{equation}
(cf. \citealt[Section 4.7]{roy88}). Here $b$ is the impact parameter, $e$ is the eccentricity ($e>1$ for a hyperbola), and  $H$ is the hyperbolic anomaly. The unit vector $\vhat{x}$ points from focus to periapse, and $\vhat{y}$ is the unit vector perpendicular to $\vhat{x}$ in the plane of motion, oriented such that $\vhat{x}\cross\vhat{y}$ is in the direction of the orbital angular momentum (see Fig.~\ref{fig:orbit-diagram}).

The eccentricity, $e$, is related to the asymptotic velocity, $v$, of the perturber relative to the pulsar, by
\begin{equation}
e=\sqrt{1+\paren[Big]{\frac{bv^2}{GM}}^2},
\end{equation}
where $G$ is the universal gravitational constant and $M$ is the total mass of the pulsar and the perturber. All the perturbers considered here have masses much less than $1\,M_\sun$, so, to a good approximation, $M$ is equal to the mass of the pulsar (or, in the Earth-term case, the Sun). The asymptotic velocity, $v$, appearing here is also called the hyperbolic excess velocity. The hyperbolic anomaly, $H$, is related to time, $t$, by the hyperbolic Kepler equation:
\begin{equation}\label{eqn:kepler}
t=t_0+\frac{b\mkern1mu\paren{e\sinh{H}-H}}{v\sqrt{e^2-1}}.
\end{equation}
Here $t_0$ is the time of periapsis, at which $H=0$. Importantly, as long as $e>1$, $t$ is an increasing function of $H$, and equation~(\ref{eqn:kepler}) may be inverted to give $H$ as a function of $t$.

The pulsar's position, $\vec{x}$, relative to the solar system barycenter is given by
\begin{equation}\label{eqn:abs-pos}
    \vec{x}\of{t}=\vec{\bar{x}}\of{t}-\frac{m}{M}\vec{r}\of{t},
\end{equation}
where
\begin{equation}
\vec{\bar{x}}\of{t}=\vec{\bar{x}}_0+\vec{\bar{v}}t
\end{equation}
is the unperturbed trajectory of the pulsar, which coincides with the trajectory of the pulsar-perturber center of mass when a perturber is present. Letting 
\begin{equation}
\Delta\vec{x}\of{t}=\vec{x}\of{t}-\vec{\bar{x}}\of{t}
\end{equation}
be the perturbation to the pulsar's position, and using~(\ref{eqn:hyp}) for $\vec{r}$, we find
\begin{equation}\label{eqn:displacement}
\Delta\vec{x}\of{t}=-\frac{mb}{M}\paren{\frac{e-\cosh{H\of{t}}}{\sqrt{e^2-1}}\mkern1mu\vhat{x}+\sinh{H\of{t}}\mkern1mu\vhat{y}}.
\end{equation}

Perturbations to the times of arrival of pulses are caused by changes in the length $d=\abs{\vec{x}}$ of the path from the pulsar to the solar system barycenter. Let $d_0=\abs{\vec{x}_0}$ be the path length at $t=0$ and $\vhat{n}$ be the unit vector pointing from the position of the pulsar at $t=0$ to the solar system barycenter, so that $\vec{\bar{x}}_0=-d_0\vhat{n}$. We then have
\begin{equation}
d=d_0\sqrt{1-\frac{2\mkern1mu\Delta\vec{x}\cdot\vhat{n}}{d_0}+\frac{\abs{\Delta\vec{x}}^2}{d_0^2}}.
\end{equation}
For $d_0\gg\abs{\Delta\vec{x}}$, $d$ can be expanded in powers of $\norm{\Delta\vec{x}}/d_0$:
\begin{equation}\label{eqn:d-expand}
d\approx d_0-\Delta\vec{x}\cdot\vhat{n}+\frac{\norm{\Delta\vec{x}}^2-\paren{\Delta\vec{x}\cdot\vhat{n}}^2}{2d_0}.
\end{equation}
The successive terms in equation~(\ref{eqn:d-expand}) have, respectively, constant, dipolar, and quadrupolar dependences on $\vec{x}$. Since the constant term contributes only to the absolute pulse phase, the TOA perturbations are given by
\begin{equation}\label{eqn:delta-tau-expand}
\Delta\tau=\frac{d-d_0}c\approx-\frac{\Delta\vec{x}\cdot\vhat{n}}{c}+\frac{\norm{\Delta\vec{x}}^2-\paren{\Delta\vec{x}\cdot\vhat{n}}^2}{2\mkern1mu c\mkern1mu d_0}.
\end{equation}
Using equations~(\ref{eqn:abs-pos})--(\ref{eqn:displacement}), this becomes
\begin{equation}\label{eqn:delta-tau-contrib}
\Delta\tau\approx -\frac{\vec{\bar{v}}\cdot\vhat{n}}{c}\mkern2mu t + \frac{\abs{\vec{\bar{v}}\times\vhat{n}}^2}{2\mkern1mu c\mkern1mu d_0}\mkern1mu t^2 + \frac{m\vec{r}\cdot\vhat{n}}{Mc},
\end{equation}
where terms involving products of the small quantities $\abs{\vec{r}}/d_0$ and $\abs{\vec{\bar{v}}}/c$ have been dropped. The remaining terms represent, respectively, the integrated Doppler shift of the pulsar's spin period; the Shklovskii correction to the period derivative \citep{shklovskii70}; and the delay caused by the reflex motion of the pulsar. From this point forward, we will ignore the Doppler and Shklovskii terms, as the low-order polynomial effects they introduce are degenerate with the period and intrinsic spin-down rate of the pulsar, and focus only on the perturbation due to reflex motion:
\begin{equation}\label{eqn:delta-tau-reflex}
    \Delta\tau=\frac{m\vec{r}\cdot\vhat{n}}{Mc}.
\end{equation}

The orientation of the interaction relative to the observer can be characterized by two angles (see Fig.~\ref{fig:orbit-diagram}) --- the inclination, $i$, of the orbital plane with respect to the plane of the sky, and the argument of periapse, $\omega$, the angle between the line of nodes and the line of apsides. A third angle, the longitude of the ascending node, $\Omega$, gives the orientation of the line of apsides relative to a reference direction, and is necessary to fully specify the orientation of the orbit with respect to a fixed coordinate system. However, it does not affect the distance between the observer and any point on the orbit, so it is irrelevant to the size of the timing perturbations under consideration.

We use the convention that $\omega$ is measured in the direction of motion and is positive when the angle from the line of apsides to the line of nodes, in the direction of motion, is acute. Furthermore, $i$ is positive when the periapse of the asteroid is closer to the observer than the center of mass. Subject to these conventions, the unit vector in the direction of the line of sight is given by
\begin{equation}
\vhat{n}=\sin{i}\sin{\omega}\,\vhat{x}+\sin{i}\cos{\omega}\,\vhat{y}+\cos{i}\,\vhat{z}.
\end{equation}
The TOA perturbation then becomes
\begin{equation}\begin{split}\label{eqn:delta-tau}
\Delta\tau=\frac{mb}{Mc}\sin{i}\paren{\frac{e-\cosh{H}}{\sqrt{e^2-1}}\mkern1mu\sin{\omega}+\sinh{H}\cos{\omega}}.
\end{split}\end{equation}
Together with the relation between hyperbolic anomaly, $H$, and time, $t$, given by equation~(\ref{eqn:kepler}), this completely specifies the form of the timing perturbations produced by a close encounter. 

\citet{drtz19} also arrived at an expression, analogous to equation~(\ref{eqn:delta-tau}), giving the expected timing signal resulting from a close encounter between a pulsar and a pointlike perturber. Our result differs in two key respects. First, while the \citeauthor{drtz19} result relies on the assumption that the orbit is highly unbound ($e\gg 1$), equation~(\ref{eqn:delta-tau}) has no such limitation, and is exact for all unbound orbits. Second, whereas \citeauthor{drtz19} expressed their result as a fractional change in frequency, our result is written in terms of TOA delays that are directly measurable. 
To compare our expression with the \citeauthor{drtz19} result, differentiate equation~(\ref{eqn:delta-tau}) with respect to $t$ and expand to first order in the quantity
\begin{equation}
\frac{GM}{bv^2}=\frac1{\sqrt{e^2-1}},
\end{equation}
which is small when $e\gg 1$. This gives
\begin{equation}\begin{split}
\der{\Delta\tau}{t}=\frac{\Delta\nu}{\nu}&\approx\frac{Gm}{bvc}\frac{\paren{\cos\omega-\sinh{H}\sin\omega}}{\cosh{H}}\\
&\qquad+\frac{mv}{Mc}\sin{i}\cos\omega,
\end{split}\end{equation}
which is equivalent (up to an additive constant that is degenerate with the unperturbed velocity of the pulsar) to equation (8) in \citet{drtz19}.

\begin{figure*}
    \centering
    \includegraphics[width=0.8\linewidth]{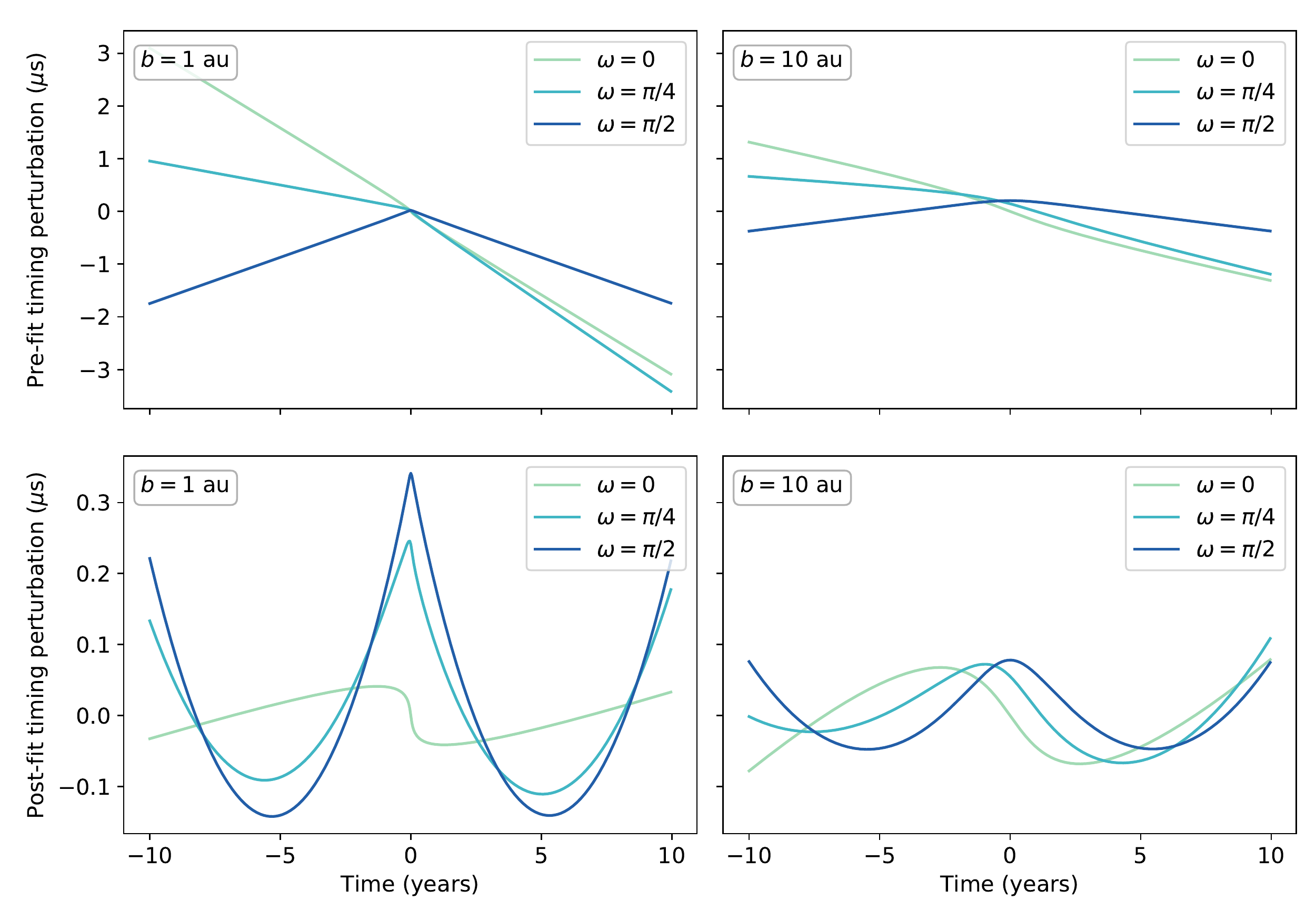}
    \caption{Timing perturbations caused by gravitational encounters with several different geometries. The upper two panels show the signal shape before fitting, while the lower two show the same signals after the model with the best-fit period and period derivative has been subtracted. All are shown for a putative 20-year data set, and correspond to an eccentricity $e=2$, a pulsar mass $M=1.4\,M_\sun$, and a projected perturber mass $m\sin{i}=10^{-10}\,M_\sun$. The left panels correspond to an impact parameter $b=\SI{1}{au}$, while the right panels correspond to $b=\SI{10}{au}$. The angle $\omega$ is the argument of the periapsis, with $\omega=\pi/2$ corresponding to a case in which the periapsis occurs along the line of sight between the Earth and the pulsar.}
    \label{fig:cmp_eg}
\end{figure*}

\subsection{Timing signature}

Over the course of an encounter between a pulsar and a perturbing object, the slope of the TOA delay changes from its initial value to a different, stable final value. This is because momentum has been transferred from the perturber to the pulsar, and so the projection of the pulsar's velocity onto the line of sight has changed. The long-term change in the pulsar's velocity is given by
\begin{equation}\label{eqn:delta-v}
\Delta\vec{v}_p=\frac{2\mkern1mu m v_0}{Me}\,\vhat{x},
\end{equation}
which corresponds to a change in the slope of the delay of
\begin{equation}\label{eqn:delta-toa-dot}
\Delta\dot\tau=-\frac{\Delta\vec{v}_p\cdot\vhat{n}}{c}=-\frac{2\mkern1mu mv_0}{Mc\mkern2mu e}\sin{i}\sin\omega.
\end{equation}
Notice that this can be either positive or negative, depending on the signs of $i$ and $\omega$. This permanent change in the velocity of the pulsar, and hence in the slope of the timing residuals, means that a long data set including an encounter will show a distinct change in apparent spin frequency.


Fig.~\ref{fig:cmp_eg} shows the shape of the timing perturbations for several particular geometries: the antisymmetric $\omega=0$ and symmetric $\omega=\pi/2$ cases, as well as the intermediate $\omega=\pi/4$ case, for impact parameters of \SI{1}{au} and \SI{10}{au}. The signals are shown both before (upper panels) and after (lower panels) fitting and removing a quadratic model describing the period and period derivative of the pulsar (see Section~\ref{sec:pulsar-encounters} below). In the post-fit perturbation shown in the lower panels, the maximum amplitude of the signal is substantially smaller, and the change from one asymptotic slope to another is obscured. Nonetheless, the post-fit perturbation for a $10^{-10}\,M_\sun$ perturber can be hundreds of nanoseconds, comparable to the signal expected from the gravitational wave  stochastic background. Comparing the \SI{1}{au} (right) and \SI{10}{au} (left) cases shows that the difference between favorable and unfavorable geometries is more significant for close encounters than it is for glancing encounters.

\subsection{Similar signals}
The timing signal produced by a close encounter superficially resembles a glitch, an abrupt change in the period of a pulsar thought to be caused by an interaction between the crust and the superfluid interior. Glitches are mainly observed in canonical pulsars, with only two cases seen so far in millisecond pulsars~\citep{cb04,mjs+16}. Glitch signals differ from those produced by close encounters in that they may be accompanied by changes in period derivative, sometimes display exponential recovery to the previous state, and have a preferred direction, spinning up the pulsar rather than slowing it down in all but a few anomalous cases (termed ``anti-glitches''; \citealt{akn+13}). Exponential recovery has not been identified in either of the observed millisecond pulsar glitches, but this is not surprising since the frequency of recovery has been shown to decrease as a pulsar's characteristic age increases~\citep{lps95}. Nevertheless, this means it is possible that one or both of the millisecond pulsar glitches could be misidentified gravitational scattering events. Assuming reasonable values for the hyperbolic excess velocity ($v=\SI{100}{km/s}$) and interaction timescale ($t_i=\SI{30}{days}$; see Section~\ref{sec:pulsar-encounters} below), the measured fractional changes in period of \num{9.5e-12} \citep{cb04} and \num{2.5e-12} \citep{mjs+16} correspond to perturber masses of $\num{2.0e-7}\,M_\sun$ ($0.067\,M_\earth$) and $\num{5.2e-8}\,M_\sun$ ($0.017\,M_\earth$), respectively.

The same kind of ramp-like signal, involving a persistent change in period, could also be produced by a gravitational wave burst with memory (BWM; \citealt{vhl10,cj12}; \citealt{mcc17}). An Earth-term BWM is distinguishable from an Earth-term gravitational scattering event in that the BWM produces quadrupolar correlations between pulsars, whereas the gravitational scattering event produces dipolar correlations. A pulsar-term BWM, however, is almost indistinguishable from a pulsar-term gravitational scattering event, except perhaps by its amplitude and time scale, and the detailed shape near the center of the event (when it is resolvable). Compared to BWMs, gravitational scattering events are much more likely to produce a gradually varying (rather than cusp-like) signal, since the interaction timescale can easily be many years, or even longer than the observing span. On the other hand, the BWM signals produced by mergers of binary black holes with astrophysically realistic masses will always be cusp-like. Some BWM events may have electromagnetic counterparts, which would serve to distinguish them from gravitational scattering events. However, any searches for such counterparts will be complicated by the fact that BWM signals may not be detected until years after the merger events that produce them.

\section{Detecting pulsar encounters}\label{sec:pulsar-encounters}

For the timing perturbation produced by an encounter between an ISO and a pulsar to be detectable, it must have an amplitude large enough to be distinguishable from noise. An important consideration here is that ISO encounter signals are always at least partially degenerate with terms involving the period and period derivative of the pulsar, and so it is appropriate to measure the amplitude only after those terms have been removed. The lower panels of  Fig.~\ref{fig:cmp_eg} show the effect of removing the period and period derivative terms in a few sample cases, illustrating that this can reduce the signal amplitude significantly.

Searching for an ISO encounter signal in a time series of TOA residuals involves fitting a model of the form
\begin{equation}
    \Delta\tau_i = a x_i + \eps_i.
\end{equation}
Here $\Delta\tau_i$ is the TOA residual at epoch $i$, $a$ is the amplitude of the signal, $x_i$ is its shape (normalized to unit amplitude), and $\eps_i$ is the noise. The least-squares estimate of the amplitude parameter is
\begin{equation}
    \hat{a}=\frac{x_i\Sigma^{-1}_{ij}\Delta\tau_j}{x_i\Sigma^{-1}_{ij}x_j},
\end{equation}
where $\Sigma_{ij}^{-1}$ is the inverse covariance matrix of the noise, and there is an implied sum over each pair of repeated indices. The corresponding uncertainty is
\begin{equation}
    \sigma_a = \paren{x_i\Sigma^{-1}_{ij}x_j}^{-1/2}.
\end{equation}
A signal is said to be detected if $z=\hat{a}/\sigma_a$ exceeds a given threshold, $z_0$, which may be set to obtain a particular false positive probability, using the fact that $z$ follows a standard normal distribution when the true amplitude, $a$, is zero. A reasonable default is $z_0=3$, but for searches over a large number of test shapes, larger values of $z_0$ may be necessary. It follows that for a signal to be detectable, its amplitude should satisfy $a\gtrsim z_0\sigma_a$, or
\begin{equation}\label{eqn:detect-criterion}
    \paren{\Delta\hat\tau_i\Sigma^{-1}_{ij}\Delta\hat\tau_j}^{-1/2} \gtrsim z_0,
\end{equation}
where $\Delta\hat\tau_i=ax_i$ is the model TOA residual at epoch $i$. When the timing noise is uncorrelated, which holds for radiometer and jitter noise, but not for scintillation or spin noise, we have $\Sigma^{-1}_{ij}=\sigma_\tau^{-2}\delta_{ij}$, where $\sigma_\tau$ is the timing precision, so equation~(\ref{eqn:detect-criterion}) reduces to the simpler form
\begin{equation}
    \Delta\hat\tau_{\mr{rms}}\gtrsim\frac{z_0\mkern2mu\sigma_\tau}{\sqrt{N}}.
\end{equation}
Here $\Delta\hat\tau_{\mr{rms}}=\paren{\Delta\hat\tau_i\mkern2mu\Delta\hat\tau_i}^{1/2}$ is the RMS of the model TOA residuals, and $N$ is the number of data points (epochs).

\begin{figure}
    \centering
    \includegraphics[width=\linewidth]{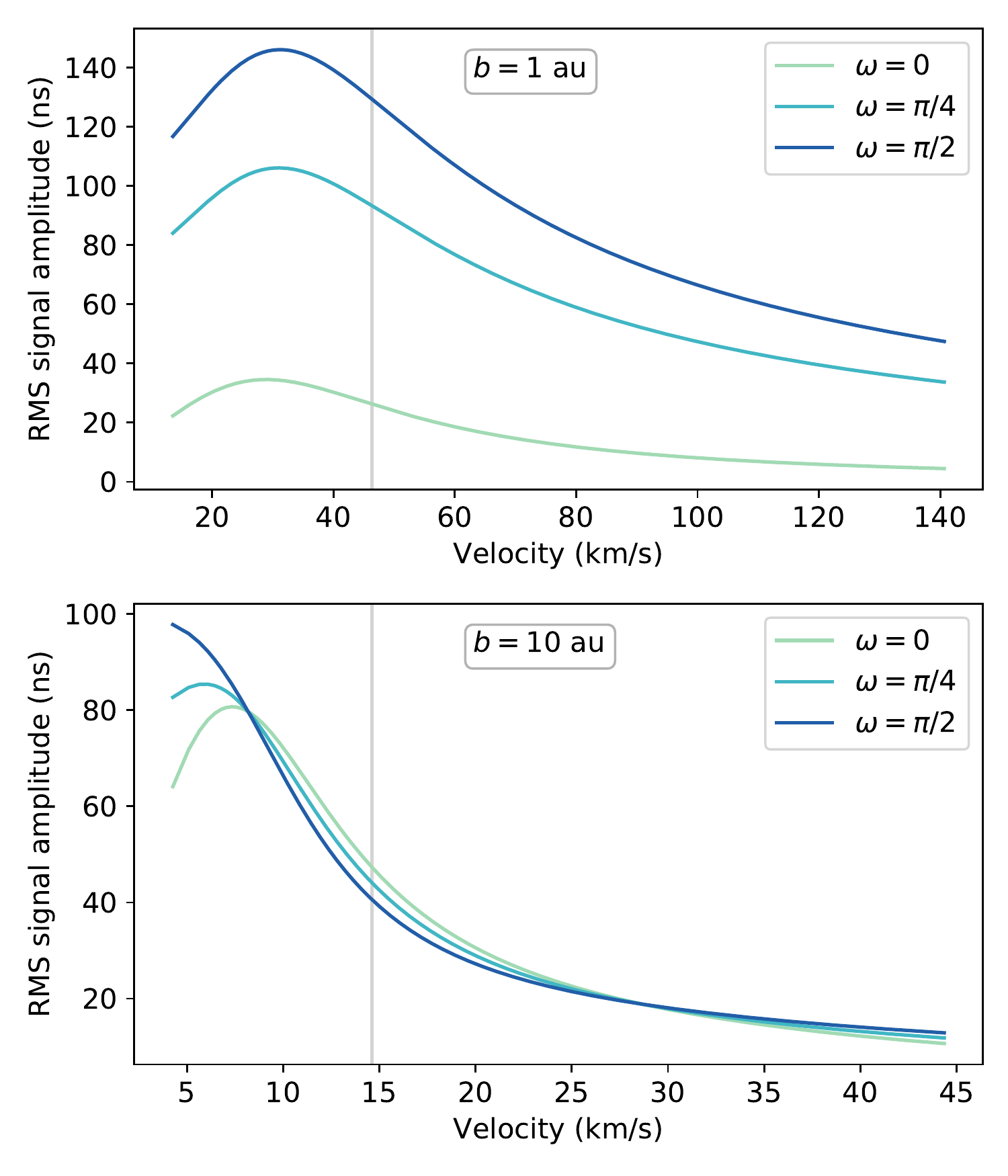}
    \caption{The RMS signal amplitude, after removing the best-fit period and period derivative, in a fiducial 20-year pulsar timing data set, as a function of hyperbolic excess velocity $v$. The top panel corresponds to encounters with an impact parameter $b=\SI{1}{au}$ and the bottom panel corresponds to $b=\SI{10}{au}$. The three lines in each panel correspond to different values of $\omega$. The amplitudes are for $m\sin{i}=10^{-10}\,M_\sun$ and scale linearly with the mass, $m$, of the perturber. As in Fig.~\ref{fig:cmp_eg}, a fiducial pulsar mass $M=1.4\,M_\sun$ has been used. The light gray vertical line indicates the velocity also used in Fig.~\ref{fig:snrshift}. In the $b=\SI{1}{au}$ case, the interaction timescale is much shorter, so the bulk of the signal comes from the difference in the pulsar's line-of-sight velocity before and after the encounter. This means that for $\omega=0$, where the net transfer of momentum is perpendicular to the line of sight, the signal amplitude is much smaller.}
    \label{fig:snr}
\end{figure}

\begin{figure}
    \centering
    \includegraphics[width=\linewidth]{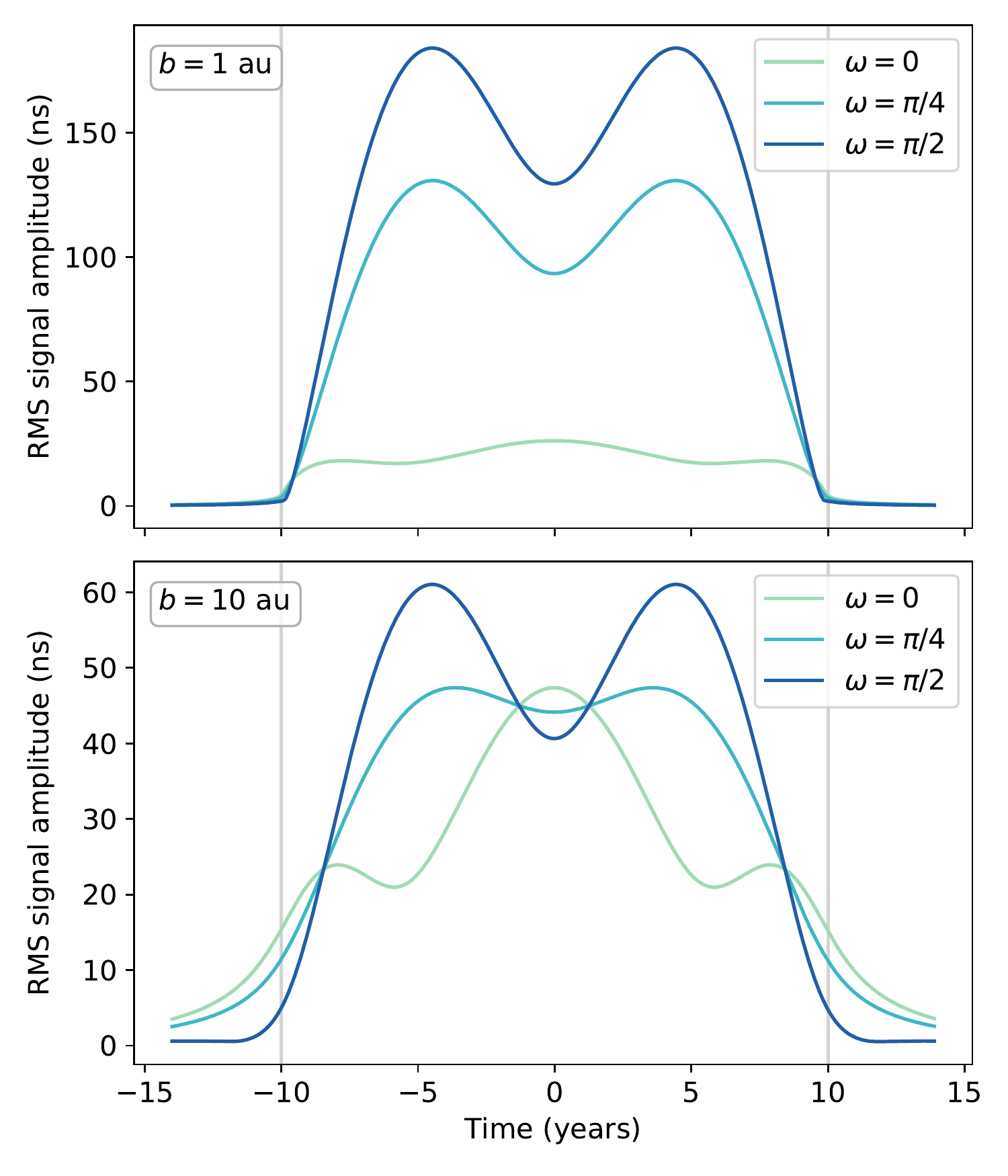}
    \caption{The RMS signal amplitude in a fiducial 20-year pulsar timing data set as a function of time of periapsis, measured from the center of the data set. The top panel corresponds to encounters with an impact parameter $b=\SI{1}{au}$ and the bottom panel corresponds to $b=\SI{10}{au}$. The three curves in each panel correspond to different values of the argument of periapsis $\omega$}, and the light gray lines indicate the boundaries of the data set. As in Fig.~\ref{fig:cmp_eg}, the curves are for $M=1.4\,M_\sun$, $m\sin{i}=10^{-10}\,M_\sun$, and $e=2$. The corresponding velocity in each case is indicated by the light gray vertical line in the appropriate panel of Fig.~\ref{fig:snr}.
    \label{fig:snrshift}
\end{figure}

\begin{figure}
    \centering
    \includegraphics[width=\linewidth]{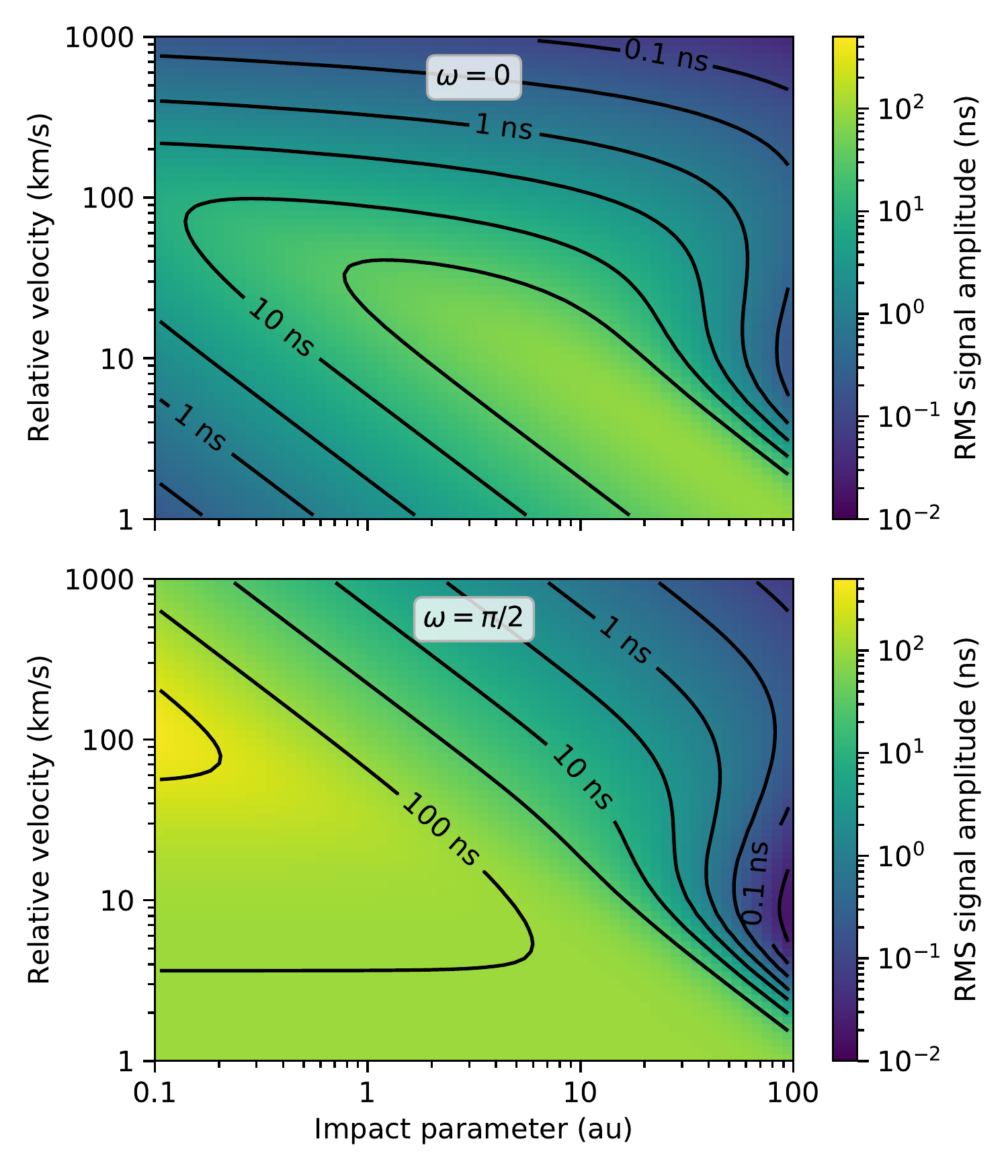}
    \caption{The RMS signal amplitude as a function of impact parameter and relative velocity, for a periapsis time of zero (corresponding to periapsis occurring at the center of the data set). As in previous figures, the amplitudes correspond to an object of mass $m=10^{-10}\,M_\sun$ with zero inclination, and a fiducial 20-year data set. The upper panel corresponds to the case $\omega=0$, where the line of apsides lies in the plane of the sky, whereas the lower panel corresponds to the case $\omega=\pi$, where the line of apsides lies along the line of sight.}
    \label{fig:vinf-b}
\end{figure}

This RMS TOA residual is plotted as a function of hyperbolic excess velocity, $v$, in Fig.~\ref{fig:snr}, and as a function of the position of the event within the data set in Fig.~\ref{fig:snrshift}. Fig.~\ref{fig:vinf-b} shows the RMS TOA residual in the two-dimensional space of impact parameter, $b$, and hyperbolic excess velocity, $v$. An important consideration in determining the detectability of a given signal is the fact that the fit for the period and period derivative of the pulsar will remove some power from the signal. For instance, interactions which take place entirely outside the data set result in a permanent velocity change, but this is degenerate with the initial velocity of the pulsar.

\begin{figure}
    \centering
    \includegraphics[width=\linewidth]{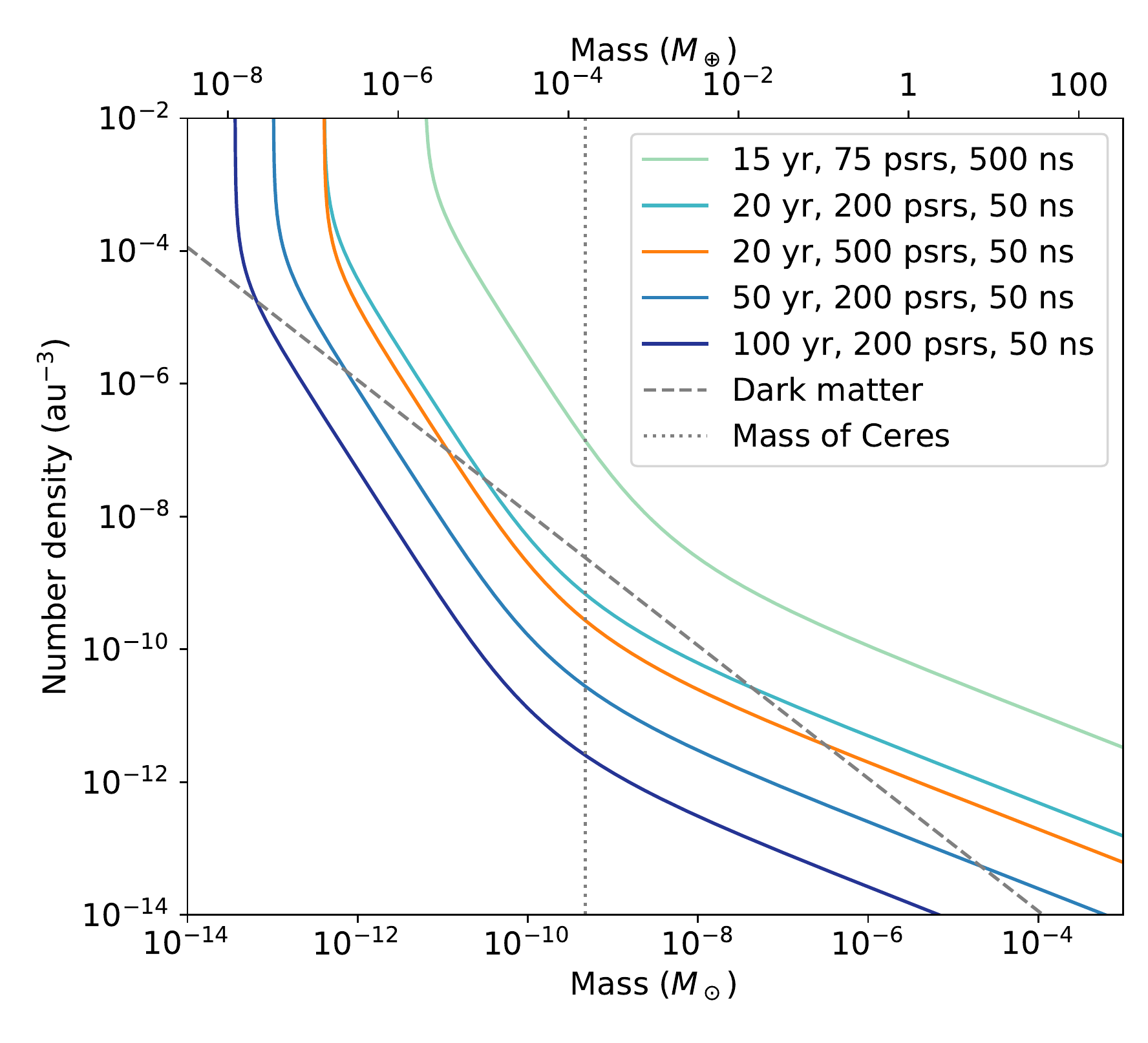}
    \caption{Detectability of interstellar populations of massive objects with a given number density and per-object mass in various PTA scenarios. For a population of objects to be detectable by a given PTA configuration, it should lie above the corresponding curve. The first case, with an observing baseline of 15 years and 75 pulsars and an RMS timing error of 500 ns, reflects capabilities similar to that of the current NANOGrav array. The remaining scenarios are increasingly optimistic future possibilities. In all cases, pulsars are assumed to be observed with a two-week cadence. The minimum detectable mass for a given number density is proportional to the timing precision. The dashed gray line indicates populations with a total mass density equal to the local density of dark matter, approximately $0.010\,M_\sun$~pc$^{-3}$.}
    \label{fig:sensitivity-curve}
\end{figure}

As seen in Fig.~\ref{fig:cmp_eg}, the shape of the signal, and thus the extent to which it is degenerate with the period and period derivative terms, depends on how the interaction time scale, $t_i\sim b/v$, compares to the data set length, $T$. In the limit where $t_i\ll T$, the signal produced by an encounter reduces to an instantaneous change in slope. This means that the signal is primarily distinguished by a sharp cusp, and the fit for the pulsar's period and period derivative does not influence its detectability, which depends only on the radial component of the the momentum exchanged during the encounter. In the opposing limit, where $t_i\gg T$, the signal is well approximated by its Taylor series expansion, and so the RMS residual falls off much faster as the event signature becomes degenerate with terms involving the period and period derivative of the pulsar. 

For interstellar objects occurring as a Poisson process with constant number density, $n$, the probability of finding $k$ objects within a volume $V$ is
\begin{equation}
    P\of{k}=\frac{\paren{nV}^k\ee^{-nV}}{k!}
\end{equation}
Assuming the same velocity, $v$, for all objects, the volume containing objects that will pass the pulsar at an impact parameter less than $b$ within the span of the data set is $V_1=\pi b^2vT$.
A PTA with $N_{\mr{psr}}$ pulsars then probes a total volume
\begin{equation}\label{eqn:volume-probed}
    V=\pi N_{\mr{psr}}V_1=\pi N_{\mr{psr}}b^2vT.
\end{equation}
It follows that the probability that a total of $k$ objects pass by the pulsars with impact parameter less than $b$ within a time $T$ is
\begin{equation}
    P\of{k}=\frac{\paren{\pi N_{\mr{psr}}nb^2vT}^k\ee^{-\pi N_{\mr{psr}}nb^2vT}}{k!}.
\end{equation}
The probability that the smallest impact parameter encountered in the data set is $b_0$ is then
\begin{equation}
    P\of{b_0} = 2\pi N_{\mr{psr}}nb_0vT\ee^{-\pi N_{\mr{psr}}nb_0^2vT}\dd b_0.
\end{equation}
The minimum impact parameter observed in a PTA data set will likely approximate the mean of this distribution, which is
\begin{equation}\label{eqn:bmin}
    \bracket{b_0}=\frac{1}{2\sqrt{N_{\mr{psr}}nvT}}.
\end{equation}
Fig.~\ref{fig:sensitivity-curve} plots the number density, $n$, against the minimum detectable mass, assuming that the signal is dominated by the event with the smallest impact parameter, and that the latter is given by equation~(\ref{eqn:bmin}).

\section{Detecting earth encounters}\label{sec:solar-system}

If an ISO passes through the solar system, it will alter the motion of the solar system barycenter in much the same way it would alter the motion of a pulsar if it passed near one. Sufficiently massive objects passing near the inner solar system, even if not directly observable, would be detectable via their influence on the orbits of the planets. For objects which are smaller or pass further from the Sun, however, this influence becomes harder to discern. One of the most sensitive ways to find these smaller or more distant objects is by using pulsar timing to measure the reflex motion of the solar system barycenter. Since the passage of such an object through the solar system perturbs the TOAs from every pulsar on the sky in a correlated way, it is possible to distinguish these flyby events from processes affecting only a single pulsar. With multiple pulsars, it is also possible to resolve signals of lower amplitude than would be detectable with a single pulsar, and to reconstruct the full geometry of the perturbing body's orbit. 

Ignoring any perturbations to the ISO's orbit caused by planets, the expected TOA signal is still that described by equation~(\ref{eqn:delta-tau}), but the geometric interpretation is slightly different. 
To compare the signals observed in different pulsars, it is useful to write the TOA perturbation in terms of the ecliptic latitude, $\beta$, and longitude, $\lambda$, of the pulsar. This can be accomplished by replacing $\vhat{n}$ in equation~(\ref{eqn:delta-tau-reflex}) with the unit vector from the solar system barycenter to the pulsar, expressed in ecliptic coordinates $\lambda$ and $\beta$:
\begin{equation}
    \vhat{n}=\sin\beta\cos\lambda\mkern2mu\vhat{u}+\sin\beta\sin\lambda\mkern2mu\vhat{v}+\cos\beta\mkern2mu\vhat{w}.
\end{equation}
Here $\vhat{u}$ is the unit vector in the direction of the vernal equinox, $\vhat{w}$ is the unit vector in the direction of the north ecliptic pole, and $\vhat{v}$ satisfies $\vhat{u}\cross\vhat{v}=\vhat{w}$. In terms of $\vhat{u}$, $\vhat{v}$, and $\vhat{w}$, the unit vectors $\vhat{x}$ and $\vhat{y}$ from section~\ref{sec:deltatau} can be written
\begin{align}
    \begin{split}
    \vhat{x} &= (\cos\omega\cos\Omega-\sin\omega\sin\Omega\cos{i})\mkern2mu\vhat{u}\\
    &\quad+(\cos\omega\sin\Omega+\sin\omega\cos\Omega\cos{i})\mkern2mu\vhat{v}\\
    &\quad+\sin\omega\sin{i}\mkern2mu\vhat{w}
    \end{split}\\
    \begin{split}
    \vhat{y} &= (-\sin\omega\cos\Omega-\cos\omega\sin\Omega\cos{i})\mkern2mu\vhat{u}\\
    &\quad+(-\sin\omega\sin\Omega+\cos\omega\cos\Omega\cos{i})\mkern2mu\vhat{v}\\
    &\quad+\cos\omega\sin{i}\mkern2mu\vhat{w}.
    \end{split}
\end{align}
The TOA perturbation for a pulsar in a direction $\vhat{n}$ is then
\begin{equation}
    \Delta\tau=\frac{mb}{Mc}\paren{\frac{ e-\cosh{H}}{\sqrt{e^2-1}}\mkern1mu\vhat{x}+\mkern1mu\sinh{H}\mkern1mu\vhat{y}}\cdot\vhat{n}.
\end{equation}
Notably, the signal has a dipolar dependence on the line-of-sight direction, $\vhat{n}$. This has a few important consequences, which are as follows: First, the signals from different pulsars differ only by a scale factor which depends in a predictable manner on the value of $\vhat{n}$ for each pulsar. This makes it possible to detect weaker signals than could be detected in a single pulsar by adding the signals from several pulsars coherently. Second, if the same event is detected in several pulsars, it should be possible to measure the orientation of the orbit completely, and thereby determine the direction the perturbing object came from. Finally, because Earth-term gravitational wave signals have a quadrupolar dependence on the line-of-sight direction, this makes Earth-term gravitational scattering signals distinguishable from Earth-term gravitational wave bursts with memory (BWMs).


Recently, errors in the solar system ephemeris have been recognized as a significant obstacle in current PTA searches for gravitational waves \citep{11yr-stochastic-background}. As a result, PTAs have developed techniques for marginalizing over uncertainties in the masses and trajectories of solar system bodies, and it has been recognized that PTA data sets can be used to make precision measurements of the position of the earth relative to the solar system barycenter. Such measurements were carried out by \citet{ipta-solar-system}, who used the first IPTA data release \citep{ipta-dr1} to constrain the masses and orbits of all the major planets and the largest five main-belt asteroids. \citeauthor{ipta-solar-system} also conducted a search for unmodeled objects in closed orbits around the solar system barycenter, using methods described in \citet{glc18}. They found no such objects larger than a few times $10^{-10}\,M_\sun$ interior to the orbit of Jupiter, with a somewhat weaker limit at larger semi-major axis values. Since their search was restricted to closed orbits, however, it did not produce direct constraints on encounters of the sort considered here.

\section{Frequency of encounters}\label{sec:freq}

The total number of asteroid-mass ISOs is currently not well understood. A weak upper bound on their number density can be derived from the assertion that such objects cannot have a total mass density exceeding the local mass density of dark matter, which is about $0.01$\kern0.25em$M_\sun$\,pc$^{-3}$ --- recent estimates range from $0.008$\kern0.25em$M_\sun$\,pc$^{-3}$ \citep{ehrn19} to $0.016$\kern0.25em$M_\sun$\,pc$^{-3}$ \citep{bcsf19}. This shows that the number density of ISOs in the solar neighborhood with masses greater than $10^{-10}\mkern2mu M_\sun$ (which is approximately 20\% of the mass of Ceres) cannot exceed approximately \SI{1.1e-8}{au^{-3}}.

The current best constraints on the abundance of these objects come from two directions. First, microlensing surveys, 
particularly the Subaru/HSC Andromeda survey \citep{nty+19}, have placed upper bounds on the frequency of large objects. Second, the detections of `Oumuamua by the Pan-STARRS1 survey~\citep{mwm+17} and C/2019 Q4~\citep{gpd+19} show that objects around \SI{1}{km} in diameter are relatively common. These detections, along with the nondetection of other similar objects by solar system surveys, can be used to constrain the number density of asteroid- and comet-like bodies in interstellar space \citep{ejv+17,dtt18}.

\citet{nty+19} report on a microlensing survey of $10^8$ stars in the Andromeda galaxy, conducted using the Hyper Suprime Cam (HSC) on the 8.2-meter Subaru telescope. The survey, which focused on constraining the abundance of PBHs, produced a single candidate microlensing event. They place an upper bound of approximately \num{2e-3} on the fraction of dark matter in the galactic halo which consists of $10^{-10}\mkern2mu M_\sun$ objects. This corresponds to a number density upper bound of about \SI{2.2e-11}{au^{-3}}.

On the solar system side, \citet{ejv+17}, writing before the discovery of `Oumuamua, find the largest number densities of asteroid- and comet-like compatible with non-detections by the Pan-STARRS1 \citep{panstarrs} and Catalina Sky Surveys \citep{clb+12}. For objects more than 1 km in diameter, their limits are \SI{1.4e-4}{au^{-3}} for comet-like ISOs, and \SI{2.4e-2}{au^{-3}} for asteroid-like ISOs. Incorporating `Oumuamua, \citet{dtt18} find a number density of \SI{0.2}{au^{-3}} for objects at least \SI{100}{m} in diameter. This can be extrapolated to objects of larger radii by assuming a power-law distribution of masses, in which the number density of objects with radii between $r$ and $r+\dd r$ is given by
\begin{equation}\label{eqn:power-law-distribution}
    \dd n \propto r^{-(\alpha+1)}\dd r.
\end{equation}

\citet{dohnanyi69} made a theoretical argument for taking $\alpha=2.5$ for material in collisional equilibrium in protoplanetary disks, and this appears to be broadly representative of small-body populations in the solar system \citep{jls02}. Attempts have also been made to determine the exponent $\alpha$ for interstellar objects empirically, based on direct measurements of interstellar dust by the \emph{Ulysses} and \emph{Galileo} spacecraft \citep{lbg+00}, and optical and radar detections of meteors identified as interstellar \citep{tbs96,baggaley00,mjm02,wb04,hajdukova08,hajdukova11,cneos140108}. On the basis of the Ulysses and Galileo data alone, \citet{lbg+00} estimated $\alpha=3.3$. Most recently, \citet{sl19-powlaw} arrived at $\alpha=3.41\pm0.17$ in an analysis incorporating `Oumuamua and the bolide meteor CNEOS 2014-01-08 \citep{cneos140108}, which is unique among meteors identified as interstellar because of its relatively large size, estimated at \SI{0.45}{m}.

Caution is necessary in applying these results to the present context, both because the power-law distribution is being extrapolated well beyond the range of sizes for which it was derived, and because the velocity measurements required to establish a meteor as interstellar in origin are challenging. \citet{hajduk01} gives a critique of the velocity measurement techniques used to establish the interstellar origin of radar meteors in the AMOR dataset \citep{tbs96,baggaley00}. Additionally, the velocity measurement used to establish the interstellar character of CNEOS 2014-01-08 relies on United States government sensors whose performance characteristics are not made public \citep{zuluaga19,dbs+19}. Nevertheless, there is currently no better way to estimate the number density of asteroid-like ISOs.

Using the \citet{dohnanyi69} scaling, the \citet{dtt18} result becomes \SI{6.3e-4}{au^{-3}} for objects larger than \SI{1}{km} and \SI{7.2e-11}{au^{-3}} for objects more than 600 km in diameter, which for a density of \SI{2.0}{g/cm^3} corresponds to a mass of about $10^{-10}\mkern2mu M_\sun$. Using the empirical scaling law from \citet{sl19-powlaw} gives considerably smaller estimates of \SI{7.8e-5}{au^{-3}} for objects larger than 1 km and \SI{2.6e-14}{au^{-3}} for objects larger than 600 km.

All of these results are estimates for the density of asteroid-mass objects in the Galaxy as a whole; it is entirely possible that local overdensities of these objects may exist, for instance in globular clusters. The density of stars near the center of a globular cluster can be hundreds to thousands of times greater  than it is in the solar neighborhood. If planetesimal-mass objects were similarly over-represented in globular clusters, it would be significantly more likely for one to interact closely with a pulsar than the previous estimates would suggest. Additionally, it is possible that populations of distant, gravitationally bound asteroids exist around at least some pulsars. Stellar flybys of the pulsar could put some of these objects on marginally unbound orbits that pass very close to the central pulsar. Such an event would be comparatively easy to detect.

\section{Summary and conclusions}\label{sec:summary}

This paper describes the shape of the pulsar timing signal that would be produced by an interstellar asteroid or other massive object in the course of flying by a pulsar or the solar system, and evaluates the likelihood of detecting such a signal in current or future PTA data sets. We find that the signal produced by a scattering event of this form would be ramp-like, since the interaction would cause a persistent change to the velocity of the pulsar relative to the solar system barycenter that would create a corresponding persistent change in the slope of the timing residuals. This is similar to the shape of the signal produced by a pulsar glitch or a gravitational wave burst with memory. The persistent nature of the signal allows SNR to build up over the course of an observing span which is long compared to the duration of the interaction.

Interactions in which a $10^{-10}\,M_\sun$ ISO passes within a few \si{au} of a pulsar or the solar system barycenter should be strong enough to be detectable with current PTAs. However, estimates of the interstellar number density of objects of this mass suggest that such interactions should be rare. No such encounters have yet been detected unambiguously.  It is possible, although unlikely, that one or both of the known millisecond pulsar glitches may be misidentified encounters with ISOs.

Current PTAs are unlikely to detect interactions between pulsars and ISOs unless the number density of ISOs is enhanced in the immediate vicinity of one or more of the pulsars, as may be the case in globular clusters. On the other hand, future PTAs with 200 or more well-timed pulsars and observing baselines of at least 20 years may be able to place astrophysically interesting constraints on compact objects with masses between $10^{-12}$ and $10^{-10}\,M_\sun$ as constituents of dark matter. The sensitivity of a PTA to ISO encounters is primarily determined by the observing span, a fact which demonstrates the benefits of timing pulsars continuously for many decades.

Because ISO encounters produce signals very similar to those expected from gravitational wave bursts with memory, candidate BWM events detected by PTAs should be carefully scrutinized to be sure they are not mistakenly identified ISO encounter signals. For Earth-term events which are detected in multiple pulsars, the pattern of spatial correlations between pulsars can distinguish between the two types of signal, but pulsar-term events will be much more challenging to interpret. The absence of a sharp cusp in the timing signal can rule out a BWM explanation in favor an ISO encounter, but the reverse is not true, as close encounters between pulsars and ISOs can produce arbitrarily sharp cusps in the timing residuals. Unless an electromagnetic counterpart can be found in archival observations, it may be impossible to distinguish a pulsar-term BWM from an ISO encounter.

\acknowledgments
The authors are members of the NANOGrav Physics Frontiers Center, which receives support from the National Science Foundation (NSF) under award number 1430284.

\bibliographystyle{aasjournal}
\bibliography{hyperbolic-orbits}

\end{document}